# Competition between Thickness and Electrical Conditioning Influence in Enhancing Giant Magnetoresistance Ratio for NiCoFe/Alq$_3$/NiCoFe Spin Valve

Mitra Djamal[1], Ramli[2], Sparisoma Viridi[3], and Khairurrijal[4]

[1]Theoretical High Energy Physics and Instrumentation Research Division,
Institut Teknologi Bandung, Jalan Ganesha 10, Bandung 40132, Indonesia
[2]Physics Department, Universitas Negeri Padang, Padang, Indonesia
[3]Nuclear Physics and Biophysics Research Division,
Institut Teknologi Bandung, Jalan Ganesha 10, Bandung 40132, Indonesia
[4]Electronic Materials Physics Research Division,
Institut Teknologi Bandung, Jalan Ganesha 10, Bandung 40132, Indonesia
*mitra@fi.itb.ac.id

## Abstract

Spacer thickness and electrical conditioning have their own influence in enhancing giant magnetoresistance (GMR) ratio. At some condition one factor can override the other as reported by experiment results. An empiric model about competition about these two factors is discussed in this work. Comparison from experiment results to validate the model are also shown and explained. A formulation is proposed to extend the existing one that now accommodates both spacer thickness and electrical conditioning in one form.

Keywords: giant magnetoresistance, thickness, electrical conditioning, model

PACS: 75.47.De, 68.55.jd

## Introduction

As tunneling magnetoresistance (TMR) [1] and giant magnetoresistance (GMR) [2] are discovered in metallic spin valves (SVs) and magnetic tunnel junctions (MTJ), widespread applications in magnetic recording and memory have been rapidly enhanced. And as sensor materials, GMR material promises some important applications, which has many attractive features, for example: low price as compared to other magnetic sensors, its electric and magnetic properties can be varied in very wide range, low-power consumption, and reduction size [3, 4]. Many factors have play important roles in enhancing GMR ratio such as impurities in spacer layer [5], electrical conditioning [6], spin-transfer torque [7], interparticle interaction [8], and spacer thickness [9-12]. Influence of the latest factor is shown in Figure 1.

Modification of model proposed by Xiong *et al.* [9] accompanied by electrical conditioning result reported by Niedermeier *et al.* [6] will be the focus in this work, in order to explain our previous results [12].

## NiCoFe/Alq$_3$/NiCoFe thin film

Thin film of NiCoFe/Alq$_3$/NiCoFe, which later acts as spin valve, has been growth at the Laboratory for Electronic Material Physics, Department of Physics, Institut Teknologi Bandung using dc-Opposed Target Magnetron Sputtering (dc-OTMS) method [12]. NiCoFe as ferromagnetic material and Alq$_3$ {(tris-(8-hydroxyquinoline) aluminum} as an organic material. are the sputtering target. Both targets are made using solid reaction. The first target is reacted with a molar ratio Ni:Co:Fe = 60:30:10, while the second is from Alq$_3$ powder. The NiCoFe/Alq3/NiCoFe thin film was grown onto Si (100) substrate.

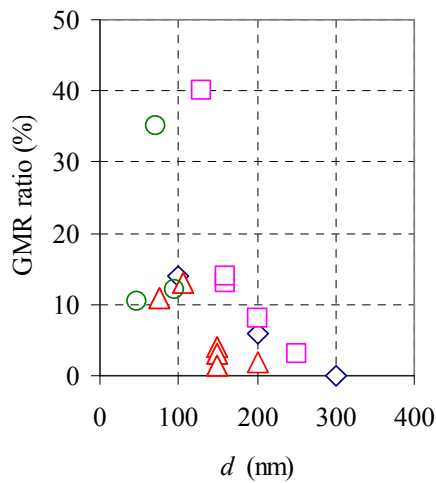

Figure 1. Spacer thickness has influence on GMR ratio as measured by: Morley *et al.* (△), Xiong *et al.* (□), Dediu *et al.* (◊), Djamal *et al.* (○).

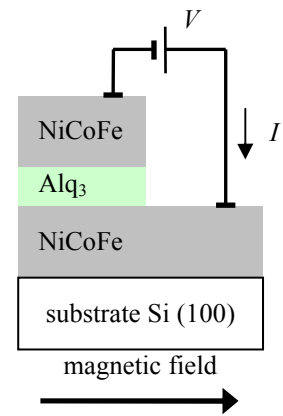

Figure 2. Schematic of NiCoFe/Alq$_3$/NiCoFe thin film and its configuration for GMR measurement.

The NiCoFe/Alq3/NiCoFe sample were deposited in several different time of growth in order to get different spacer layer thickness. Other deposition parameters are fixed. These parameters are: flow rate of Argon gas is 100 sccm, the growth pressure is 0.52 torr, dc Voltage is 600 volt, and the temperature is 1000 °C. The samples were characterized by using SEM (Scanning Electron Microscope) type JEOL JSM-6360 LA and magnetoresistance measurements were made by using a linear four-point probe method with current-perpendicular to-plane (CPP).

Table 1. Growth time, applied voltage, layers thickness, and GMR ratio for NiCoFe/Alq3/NiCoFe thin film.

| Growth time (minutes) | Applied voltage (mV) | Layer thickness NiCoFe / $Alq_3$ / NiCoFe (nm) | GMR ratio (%) |
|---|---|---|---|
| 10 | 78.4 | 100 / 48 / 100 | 10.5 |
| 15 | 154.1 | 137 / 72 / 137 | 35.0 |
| 20 | 95.2 | 175 / 96 / 175 | 12.0 |

Overall parameters in producing the thin film and measurement results (layer thickness and GMR ratio) are as given in Table 1.

**Thickness and electrical conditioning influence**

As explain in [9] the GMR ratio $\Delta R / R$ can be formulated using sprin polarization $p$ as in

$$\frac{\Delta R}{R} = \frac{R_{AP} - R_P}{R_{AP}} = \frac{2p_1 p_2 e^{-(d-d_0)/\lambda_S}}{1 + p_1 p_2 e^{-(d-d_0)/\lambda_S}}, \quad (1)$$

where $R_{AP}$ and $R_P$ stand for $R$ in the anti-parallel and parallel magnetization configurations, respectively. The spin polarization itself is defined as

$$p_i = \frac{N_{i\uparrow}(E_F) - N_{i\downarrow}(E_F)}{N_{i\uparrow}(E_F) + N_{i\downarrow}(E_F)}, \quad (2)$$

with $N_\uparrow$ is carrier density in the majority spin state and $N_\downarrow$ is in the minority spin state. Index $i$ has value of 1 or 2 for the first and second ferromagnetic (FM) contact, respectively. The parameter $\lambda_S$ is the spin diffusion length, while $d_0$ is the thickness of an 'ill-defined' layer between the conducting (FM) and spacer layer (nonmagnetic, NM). Thickness of the spacer layer is $d$. There is also extension of Equation (1), that includes temperature $T$ influence in spin polarization $p$, spin diffusion length $\lambda_S$, and new term called spin injection efficiency $\eta$, which is [13]

$$\frac{\Delta R}{R}(T) = \frac{R_{AP}(T) - R_P(T)}{R_{AP}(T)}$$
$$= \frac{2p_1(T)\eta_1(T)p_2(T)\eta_2(T)e^{-(d-d_0)/\lambda_S(T)}}{1 + p_1(T)\eta_1(T)p_2(T)\eta_2(T)e^{-(d-d_0)/\lambda_S(T)}}. \quad (3)$$

It has been observed that spin diffusion length $\lambda_S(T)$ of organic semiconductor $Alq_3$ layer increases as temperature $T$ decreasing [13] and spin polarization $p$ is also decreasing at higher temperature $T$ [9]. The value of GMR ratio itself decreases as temperature $T$ increasing [14]. The influence of temperature to GMR ratio is out of scope of this work.

From the work of Niedermeier *et al.* [6] following information through a digitizing process can be found.

Table 2. Electrical conditioning results reported by Niedermeier *et al.* [6].

| $I_{cond}$ (mA/cm$^2$) | Figure 2 in [6] | | Pixels | |
|---|---|---|---|---|
| | Time (h) | $\Delta R / R$ (%) | x | y |
| - | 0 | 0 | 94 | 42 |
| - | 10 | -20 | 537 | 404 |
| 0 | 4.108352 | -0.11050 | 276 | 44 |
| 6 | 4.108352 | -2.54144 | 276 | 88 |
| 19 | 4.108352 | -5.74586 | 276 | 146 |
| 75 | 4.108352 | -12.3757 | 276 | 266 |
| 125 | 4.108352 | -16.8508 | 276 | 347 |
| 125 | 1.986456 | -15.8011 | 182 | 328 |
| 125 | 3.995485 | -16.7956 | 271 | 346 |
| 125 | 6.817156 | -17.5691 | 396 | 360 |
| 125 | 8.329571 | -17.7901 | 463 | 364 |
| 125 | 9.6614 | -18.0663 | 522 | 369 |

Two first rows are the reference points, which are used to get the information from Figure 2 in [6]. Using this information following fitting equation can be found

$$\frac{\Delta R}{R} \approx -0.1281 I_{cond} - 1.7168, \quad (4)$$

$$\frac{\Delta R}{R} \approx -14165 \ln(t_{cond}) - 14.831, \quad (5)$$

where Equation (4) and (5) gives $r^2 = 0.9605$ and $r^2 = 0.9992$, respectively. These two equations are drawn in solid line in Figure 3.

Equation (4) suggests a rough approximation about influence of conditioning current $I_{cond}$, which is related to applied voltage $V_{app}$, to the GMR ratio

$$\frac{\Delta R}{R} \approx c_1 V_{app} + c_2, \quad (6)$$

with $c_1$ and $c_2$ is constant to be determined.

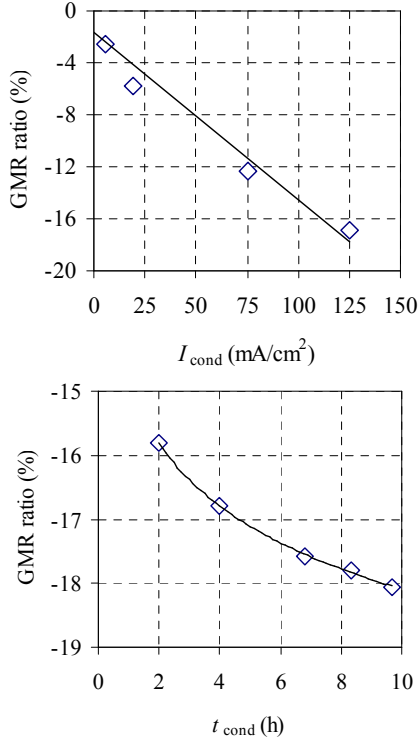

Figure 3. GMR ratio $\Delta R / R$ as function of: conditioning current $I_{cond}$ (top) and conditioning time $t_{cond}$ (bottom) as obtained from [6].

By merging Equation (6) with Equation (1) will produced a formulation

$$\frac{\Delta R}{R} = \frac{R_{AP} - R_P}{R_{AP}} = \frac{2(c_1 V_{app} + c_2) p_1 p_2 e^{-(d-d_0)/\lambda_S}}{1 + p_1 p_2 e^{-(d-d_0)/\lambda_S}}, \quad (7)$$

that accommodates the influence of both spacer or NM layer thickness and applied voltage during growth of spin valve thin film to the GMR ratio.

**Results and discussion**

Following parameters are used in Equation (7) to fit the measured result in Table 1: $c_1 = 0.96\,\text{V}^{-1}$, $c_2 = -51$, $p_1 p_2 = 0.35$, $d_0 = 11\,\text{nm}$, and $\lambda_s = 90\,\text{nm}$. The fitting result is shown in Figure 4, which gives $r^2 = 0.9959$. As it can be seen there are three different lines that correspond to each $V_{app}$, which are 78.4 V, 154.1 mV, and 95.2 mV. Then it can be explained why the measured data does not have a curve as in Figure 1 that GMR ratio is in general decreasing as spacer thickness $d$ increasing. We propose that the electrical conditioning occurred 'accidentally' in thin film growth process has enhancing the GMR ratio, which its role is dominant to the role of spacer thickness. As it can be seen, Equation (7) will be reduced to Equation (1) if the $V_{app}$ is taken to be constant in thin film growth process.

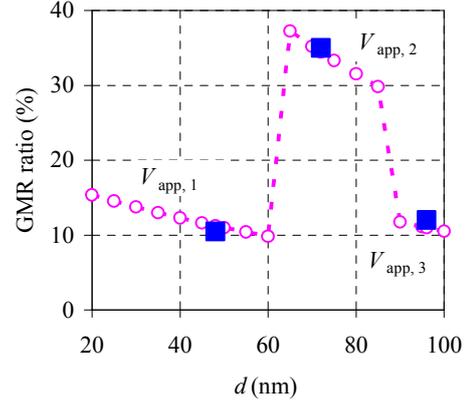

Figure 4. Plot of Equation (7) (in hollow circle, online: magenta) for measured data (filled square, online: blue) with different $V_{app}$: $V_{app,1} = 78.4\,\text{mV}$, $V_{app,2} = 154.1\,\text{mV}$, and $V_{app,3} = 95.2\,\text{mV}$

The influences of parameter $c_1$, $c_2$, $p_1 p_2$, $d_0$, and $\lambda_s$ are given in following figures.

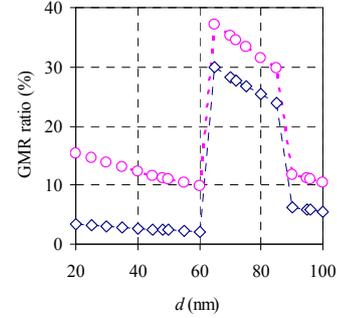

Figure 5. GMR ratio for $c_2$ with value: -51 (top curve) and -70 (bottom curve).

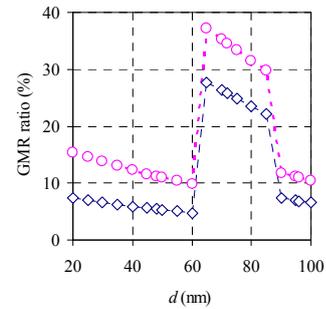

Figure 6. GMR ratio for $c_1$ with value: 0.96 V$^{-1}$ (top curve) and 0.8 V$^{-1}$ (bottom curve).

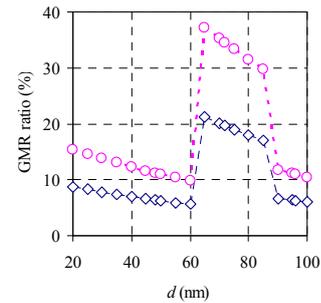

Figure 7. GMR ratio for $p_1 p_2$ with value: 0.35 (top curve) and 0.2 (bottom curve).

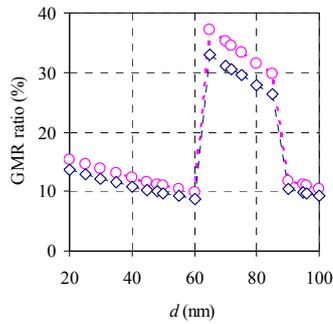

Figure 8. GMR ratio for $d_0$ with value: 11 nm (top curve) and 0 nm (bottom curve).

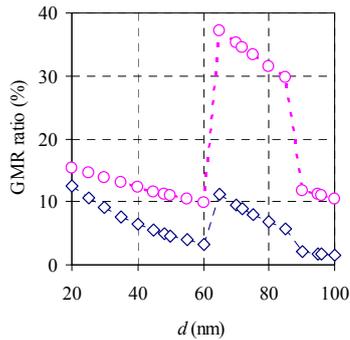

Figure 9. GMR ratio for $\lambda_s$ with value: 90 nm (top curve) and 30 nm (bottom curve).

In Figure 5 – 9, all parameters used are the same as in Figure 4 and the preceded text. Only one parameter in each figure is changed to show its influence to the curve of GMR ratio $\Delta R / R$ to spacer thickness $d$.

## Conclusion

A formulation that can accommodate both influence of spacer thickness and electrical conditioning to GMR ratio has been presented. It can explain measured result with $r^2 = 0.9959$. The influences of each parameter are also shown.

## Acknowledgements

This work was (partially) supported by the Directorate for Research and Community Service, The Ministry of National Education Republic of Indonesia through Competency Grant under contract No: 407/SP2H/PP/ DP2M/VI/2010.